\documentclass{aa}
\usepackage{epsfig}

\newcommand{\refe}{}

\begin{document}

\title{A simple model for the complex lag structure of microquasars}

\author{P. Varni\`ere}

\institute{Department of Physics \& Astronomy, Rochester University, 
Rochester NY 14627-0171
          pvarni@pas.rochester.edu}
 
\abstract{The phase lag structure between the hard and soft X-ray photons
 observed in GRS $1915$+$105$ and XTE J$1550$+$564$ has been said to be
``complex'' because the phase of the Quasi-Periodic Oscillation fundamental Fourier 
 mode changes with time
and because the even and odd harmonics signs behave differentely.  From simultaneous
X-ray and radio observations this seems to be related to the presence of a jet
(level of radio emission).  We propose a simple idea where a partial
absorption of the signal can shift the phases of the Fourier modes and account
for the phase lag reversal. We also briefly discuss a possible physical
mechanism that could lead to such an absorption of the quasi-periodic
oscillation modulation.
   \keywords{ X-rays: binaries,  
     stars: individual (GRS $1915$+$105$, XTE J$1550$-$564$), accretion disks}
}

   \maketitle
\section{Introduction}  
\label{sect:intro}

     RXTE has provided us with a better picture of the temporal
     behavior of  X-ray binaries, using such techniques as Fourier Transform (FT), {\refe time}
     lag and coherence
     computation.
     The {\refe time} lag between the low-energy ($2$ - $5$ keV) and the high-energy ($5$ - $20$ kev) is 
     generally associated with Inverse-Compton of soft photons producing hard 
     photons. {\refe Most of the time, the high energy variability lags behind the low energy
     emission ; this is the so called ``hard-lag''}.
     Surprisingly, {\refe there exist some observations where the QPO hard} lags appear to change sign 
     ({\refe becoming what is called a ``soft-lag''})
     during an observation and also between 
     observations. This is inconsistent with the Inverse-Compton explanation.
     In GRS $1915$+$105$  (e.g. Cui, 1999, Lin {  et al.}, 2000) and XTE J$1550$-$564$
     (e.g. Wijnands {  et al.}, 1999, Cui {  et al.}, 2000), this unusual time lag 
     behavior  has been reported during outburst and/or the radio-loud state. Namely, the sign of the 
     QPO's time lag changed over a single observation whereas the sign of its first harmonics'
     time lag stayed the same. 

     {\refe 
       In their 1999 paper, Wijnands {  et al.} point out that a change in the waveform of the QPO
       between the low and high energy emission could explain the sign difference in the lag of the 
       fundamental and the first harmonic, but this was not explored further.       
       Lin {  et al.} (2000)  noted that the presence of that same sign difference 
       does not imply a real time delay. For example the same effect appears 
       in the presence of a decaying oscillating signal. 
       Here we will explore in more detail what is at the origin of the fundamental lag's sign change.
       The same mechanism will also create the sign difference mentioned above without implying
       a real time delay.}

	Sect. $2$ discuss simultaneous radio and X-ray data from GRS $1915$+$105$
	taken from the plateau/hard-steady ($\chi$, Belloni {  et al.}, 2000) state \cite{M01}. 
	We  use this data to gain insight 
	into the relation between radio/jet and the X-ray timing properties of the system.
	In Sect. $3$ we show the behavior of the Fourier Transform (FT)
	in the case of an absorbed sinusoid.
	In Sect. $4$ we  make use of this simple, zeroth order, model to explain the
	complex lag structure observed in GRS $1915$+$105$ and XTE J$1550$-$564$ and see what 
	we can infer about those systems.

\section{The Case of GRS $1915$+$105$}
\label{sect:1915}

      Muno et al. (2001) studied the hard state ($\chi$ state in the classifications by 
      Belloni et al., 2000, or radio plateau) using simultaneous X-ray and radio observations. 
      {\refe In this section we will discuss Fig. $7$ and $8$ in Muno et al. (2001),
      in order to emphasize the observational constraints on the  behaviour we are trying to explain here.} 
     
      The left of their Fig. $8$
      shows how the temporal properties (QPO frequency on the top and phase lag at the
      QPO frequency at the bottom) correlate with the different components of the X-ray flux,  namely from left
      to right, the total flux, the thermal/disk flux  and the power-law flux. 
      By looking carefully at the plots two populations can be distinguished (the triangle and the cross). This
      distinction is more apparent in the graph showing the lag.

      On the left upper panel of Fig. $8$ we see that 
      for a QPO frequency higher than about two hertz, the QPO frequency 
      appears to be correlated with the total flux and the power-law flux (which in fact
      dominates the total flux). This applies for most of the low-mass X-ray binaries. 
      For a QPO frequency lower than $2$Hz, this QPO frequency no longer correlates  with
      any of the X-ray fluxes. In fact all of the frequencies below  $2$ Hz appear
      at a similar flux level for both the thermal and power-law flux, {\em .i.e} the cluster of 
      triangles is very narrow. These points are also the ones with a high radio flux (the 
      triangles represent the radio-loud state) as is seen in Fig. $7$.
      These radio-loud points are also the only ones to exhibit a positive phase lag.
      Concerning this lag, there is also another difference besides the change of sign 
      between the radio-loud  and the radio-quiet state:
      If we look at the left lower panel of Fig. $8$, there is no correlation between the lag 
      and any of the X-ray fluxes.
      However, 
      depending on wherever the source is radio-loud or radio-quiet the  ``clusters'' of points appear to be 
      perpendicular to each other.

       In the radio-loud case,  the temporal behavior of the source is modified for quasi 
	constant X-ray fluxes. 
       These modifications are a function of the radio flux.
       On  Fig. $7$ is shown the evolution of the temporal
       properties such as the QPO frequency, the phase lag, the coherence and the ratio of 
	low-frequency power as a function of the radio flux  at $15.2$ GHz.
       Once again the radio-loud and radio-quiet points are well separated. 
       The separation occurs at a radio flux of about $60$ mJy.

      By looking in more detail at the first plot (QPO frequency - radio flux) we see that a QPO frequency
       less than two hertz is always 
       associated with a radio flux of more than $60$ mJy. These same QPOs have a positive phase lag 
       and show much less coherence  than the QPOs in the radio quiet state. 
       Moreover, the phase lag which seems totally
       uncorrelated with the radio flux when it is less than $60$ mJy, appears to be correlated with 
	the higher radio fluxes. In the graph of the ratio of low-frequency power as function of the radio flux the 
	possible correlation
       seems to reverse during the transition between radio-quiet and radio-loud.

       Either these QPOs (less than $2$Hz, more than $2$Hz) arise from a different mechanism 
       ( {\em e.g.} one related with 
       the jet and the other one not) or there is a threshold in radio flux above which  new 
	phenomena appear in addition to the QPO mechanism. This could cause a  modification of 
       the temporal behavior of the source, especially relevant to the lag which seems to become
       proportional to the radio flux.
       We will focus on this last possibility. 
       The presence of two different unrelated mechanisms, {\refe one from the jet and the other  
       from the disk}, seems improbable because of the smooth 
	transition
       in QPO properties as a function of time (see for example Fig. $6$ of Muno {et al}, 2001)
       However, before exploring the possible origin for the change in temporal properties, we will
       look at the lag definition and its computation through Fourier transforms.

\section{Fourier Transform and phase lag}
\label{sect:FT}

\subsection{Definition of lag and coherence}

        We will briefly go over the definition of the lag as presented by Vaughan \& Nowak (1997).
        Suppose that $x_1(t_k)$ and $x_2(t_k)$  represent the X-ray flux in two energy bands (soft and hard)
	at time $t_k$. We note $X_1(\nu_j)$ and $X_2(\nu_j)$ as their Fourier transforms at the frequency $\nu_j$:
\begin{equation}
X(\nu_j)  = \frac{1}{2\pi} \int x(t_k) e^{-i\nu_jt_k} dt_k\nonumber
\end{equation} 

       The time lag between the hard and soft X-ray is then defined as:                    
\begin{equation}   
\delta t(\nu_j) = \frac{1}{2\pi \nu_j} \times  arg (X_1^\star(\nu_j)X_2(\nu_j))  \nonumber \\ 
                = \frac{arg(X_2) -arg(X_1)}{2\pi \nu_j} \nonumber       
\end{equation}

	Modification of one of the two phases (lowering the phase of the soft band or increasing the phase of the
	hard band) could induce the lag to change sign. More generally, any change
        in the phase of one of the bands, caused by  internal or external phenomena, could lead to a
        sign change in the lag.

	The coherence is a measure of how much of a signal $f$ can be predicted knowing a signal $h$.
	In our case it means how much of the high energy flux can be predicted knowing the
	low energy flux. If the two signals are related then the coherence is high; the maximum
	equals one, which correspond to the case where there is a linear transformation to go from one to the other.

\subsection{Lag: a simple derivation}	

        One can reproduce the observed behavior of
        the lag and harmonics using simple assumptions about the initial profile. 
	The idea is to compare the Fourier representation of an initial profile
	(here a constant plus a cosine) taken to be the hard X-ray, to a modified profile
	taken to be the soft X-ray. We will then compute the lag between them and 
	show a simple way to match the observed lag behavior.

	If we compute the Fourier Transform of a sinusoid function we obtain the frequency, amplitude
	and phase. In order to make it similar to data we take the sinusoidal profile surimposed
	on a constant background and add a small amount of random noise to it. By using the FT 
	we still find the frequency, amplitude and phase. 
	Now take into account the case where some part of the modulated emission does 
	not arrive to the
        observer but a part of it is "absorbed/obscured" by a media located in  the system.
	This would make a profile similar to the one of fig 1.

\begin{figure}[htbp]
\centering
\resizebox{\hsize}{!}{\includegraphics{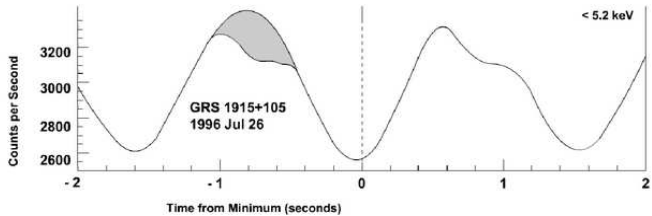}}
   \caption{{\small \cite{M97} studied GRS $1915$+$105$ timing variability using 
       average QPO-folded profile for the $0.63$Hz QPO. The profile show an absorbed-like 
       feature for the soft energy band ($< 5.2$ keV) emphasized on the figure by the grey 
       "missing part". {\refe \cite{M97} also showed the QPO profile at higher (hard) energies.
       For higher energies, the profiles appear more sinusoidal, i.e. the
       ``missing/absorb part'' becomes smaller.}
}}
\label{fig:QPO_profile}
\end{figure}
	Table $1$ shows the Fourier Transform
	using an input profile of unity plus a sinusoid with rms amplitude $rms=0.14$ 
	minus a Gaussian profile of amplitude $\gamma$ centered to
	reproduce a profile like the one from the Fig. $15$ of 
	\cite{M97}\footnote{In the first step of this work we searched for which type of profiles
	are able to reproduce the lag structure. In a second step we tried to find similar
	 profiles in observations. The paper \cite{M97} has this ``absorbed''-like profile
	we use as an example here.}. The first line is the representation of the initial
	state, the test \# 1 shows the first two  frequencies of the Fourier representation
	of the absorbed signal.

\begin{table}[htbp]
\caption{Representation obtained from the FT of a signal $1 +\cos \phi$ plus an absorption of
amplitude $\gamma$.}
\label{Tab:run1}
\centering
\begin{tabular}{ccccc}
\hline
test &  $\gamma$      & freq & amp & phase \\
  \hline
\# 0 &   $0$     & $1$  &$0.14$ & $0$      \\
\# 1 &   $0.07$  & $1$  &$0.12$ & $-0.02$  \\
     &   $0.07$  & $2$  &$0.016$& $1.27$   \\
\hline
\end{tabular}
\end{table}

	We see that by doing the FT on this signal we obtain different 
	parameters for the sinusoid. 
	Depending on the amount of "absorption" we can obtain a  
	smaller value for the amplitudes but the striking feature is the effect on
	the phase: a change is observed.
	Moreover, the sign of the phase difference is not the same for the fundamental and
	its first harmonics.
	If we take the formula for phase lag and say that only the low energy/soft X-rays
	are absorbed and not the hard ones  we can compute the phase lag which appears as
	a consequence of the absorption of only part of the signal. Doing so  reproduces
	the observed phase characteristics: a different sign for the fundamental and first harmonics.
	In addition, if the absorption is turned on,  it creates a change in the sign of the lag.
	This comes from the fact that the FT adjusts the data with a shifted sinusoid, creating 
	a phase difference. We propose that this is the origin of the changing sign of the lag presented in the 
	previous section. This will also decrease the coherence between the two bands
	as a new signal is added to only one band. This  happens without changing the  primary physical 
	phenomena that produces the emission in the two bands.
	The above results can be easily  illustrated even using two sinusoidal signals with a $\pi/2$ phase between them:
\begin{equation}   
\cos \theta + \epsilon\  \sin \theta = \frac{1}{\cos \phi}\  \cos(\theta -\phi) , 
\tan \phi = \epsilon 
\end{equation}  

         The presence of a second, small, sinusoidal signal with a phase lag of $\pi/2$ and 
	an amplitude $\epsilon$ is
	 enough to create an ``apparent'' phase lag of $\phi$= atan$(\epsilon)$, which is
	 about $\epsilon$, the amplitude of the perturbation. 

	If we now add the presence of a small harmonic 
	to the QPO (of amplitude label $rms2$ in table $2$) and compare the
	result from the FT to that with the  the same signal absorbed, the effect on the phase is
	even more striking. Table $2$ shows the results of such a simulation.
	Indeed, the induced phase lag between the real data and the absorbed one
	does not have the same sign at the fundamental vs. the first 
	harmonics. This could be at the origin of the observed phenomena.
	{\refe  In this work 
	we show that an absorption of the low energy part of the signal will give the sign difference
	in the lag and also explain the observed change of sign for the fundamental.}

\begin{table}[htbp]
\caption{Representation obtain from the FT of a signal $1 +\cos \phi + rms2\cos(2\phi)$ plus an absorption of
amplitude $\gamma$.}
\label{Tab:run2}
\centering
\begin{tabular}{ccccccc}
\hline
test &  rms2& $\gamma$      & freq & amp & phase & lag     \\
\hline
\# 2 & $0.07$&  $0.0$   & $1$  &$0.14$ & $0$     &         \\
     & $0.07$&  $0.0$   & $2$  &$0.07$ & $-\pi/2$&         \\
\# 3 & $0.07$&  $0.07$  & $1$  &$0.12$ & $-0.02$ &  0.02   \\
     & $0.07$&  $0.07$  & $2$  &$0.054$& $-1.47$ &  -0.1   \\
\# 4 & $0.07$&  $0.1$   & $1$  &$0.112$& $-0.03$ &  0.03   \\
     & $0.07$&  $0.1$   & $2$  &$0.048$& $-1.42$ &  -0.15  \\
\hline
\end{tabular}
\end{table}

\section{Application to microquasar observations}
\label{sect:application}

	Using the above argument it appears that the use of an FT can lead to an incorrect interpretation
	of the lag in the presence of an absorption which depends on the energy band.
	To use this idea for the observations of GRS $1915$+$105$ presented in Sect. $2$,  we
	need to find what may produce the "absorbed" part of the QPO modulation. 
	This has to be related to the jet,
	either having the same origin, or being a consequence of it. In the
	following we will assume that the QPO modulation
	is created by a hot spiral/point, {\refe for example in Varni\`ere et al. (2002, 2005 in preparation)
	}, and we are just interested in further 
	absorption/modulation of this already existing modulation.
	{\refe  As mentioned before, we choose to keep the same mechanism for
	   the QPO above and below $2$Hz. Another possibility is that the QPO above $2$Hz
	comes from the disk while the one below $2$Hz is coming from the jet. This however,
	seems improbable because the passage through $2$Hz is smooth in all 
	variables (see Fig. $6$ of Muno {  et al} (2001).}

	Suppose that the basis of the jet/corona gets "between" the observer and the spiral during one
	orbit of the spiral in the disk. This is enough to "absorb" a part of the flux modulation, especially
	if it happens when the spiral is "behind" the black hole and therefore near the maximum
	of the modulation. This simple model is able to explain both the occurrence of changing sign
	lag and  its
	relation with the jet. In the same way it can also explain the fact that absorption is energy 
	dependant, which makes the coherence drop. 
	In fact, anything located inside the inner radius of the disk that can absorb a small part
	of the flux coming 
	from the hot spiral could explain the changing sign of the lag and the complex behavior of
	the harmonics. But this needs to be related to the radio flux and therefore to the jet mechanism.
\newline

	The first way to check this idea is to look at the QPO profile and see if there is an energy
	dependant departure 
	from a sinusoidal signal. Morgan et al. (1997) 
	show for the low-frequency QPO that there is indeed a departure from a sinusoid, which seems
	compatible with an absorption feature. This kind of analysis is difficult and rarely done for
	QPOs because of the lack of photons at these timescales. 
	Another way to check the same properties is to see how the value of the lag depends on the
	energy band chosen. Using the idea of an energy dependant absorption we see that the 
	negative lag
	will be more important between the lower energy band (say, $2-4$keV) and the highest possible 
	band available, than between two high energy bands.
	It seems possible to have a change of the sign of the lag if we look to high enough 
	energies (for example using INTEGRAL data).
\newline

        This simple model can also be used with observational data to gain insight into
	the geometry near the black hole. The pulse shape of the QPO in different
	energy bands can allow us to constrain the relative geometry of the absorption
	region with respect to the emissive region (QPO origin), and also the column
	density of the absorber. We will test several mechanisms that could lead to
	this ``absorbed-like'' profile and compare them with observationnal data.

\section{Conclusions}
\label{sect:conclusion}

        This letter shows how absorption can modify the X-Ray signal and give rise to
	an apparent change in the phase lag between the hard and soft photons. The
	model is phenomenological, and future simulation work is needed to yield more
	quantitive predictions that can be compared with observational data and
	thereby giving access to the geometry in the inner part of the disk. 
	Indeed, with numerical simulation  we intend to probe the relative geometry 
	of the QPO emission region with respect to the absorbing media by using the shape of the QPO pulse.
	The use of RXTE and INTEGRAL data
	together with numerical simulations of the absorption of a ``hot-spot''
	orbiting in the disk will further test this idea.

\begin{acknowledgements}
PV is supported by NSF grants AST-9702484, AST-0098442, NASA
grant NAG5-8428, HST grant, DOE grant DE-FG02-00ER54600, the
Laboratory for Laser Energetics and the french GDR PCHE.

PV thanks Michel Tagger, Eric Blackman, Jason Maron, Jerome Rodriguez and Mike Muno for all the discussions, 
helpful comments on the paper and data. PV thanks the annonymous referee for the comments that helped
to clarify the paper.
\end{acknowledgements}

\end{document}